\documentclass{article}
\usepackage{spconf,amsmath,graphicx}
\usepackage{epstopdf}
\usepackage{graphicx} % For including images
\usepackage{amsmath}  % For mathematical symbols and equations
\usepackage{hyperref} % For hyperlinks
\usepackage{listings} % For including code listings
\usepackage{cite}
\usepackage{algorithmic}
\usepackage{array, makecell}
\usepackage{multicol}
\usepackage{multirow}
\usepackage{color}
\usepackage{xspace}
\usepackage{booktabs}
\usepackage{pgfplots}
\usepackage{tikz}
\usepackage{pgfplotstable}
\usepackage{caption}
\usepackage{enumitem}
\def\seta{{\bf Set A}}
\def\setb{{\bf Set B}}
\def\setc{{\bf Set C}}

\title{A cross-talk robust multichannel VAD model for multiparty agent interactions trained using synthetic re-recordings}
\name{$^{a}$Hyewon Han\thanks{This work was done during an internship at Disney Research}, $^b$Naveen Kumar}
\address{$^a$Dept. of Electrical and Electronic Engineering, Yonsei University, Seoul, South Korea\\
    $^b$Disney Research Imagineering, Los Angeles, USA}
\begin{document}
% \tenpt
\textbf{IEEE Copyright Notice}
\vspace{10pt}

© 2024 IEEE. Personal use of this material is permitted. Permission from IEEE must be obtained for all other uses, in any current or future media, including reprinting/republishing this material for advertising or promotional purposes, creating new collective works, for resale or redistribution to servers or lists, or reuse of any copyrighted component of this work in other works.
\ninept
\maketitle
\begin{abstract}
In this work, we propose a novel cross-talk rejection framework for a multi-channel multi-talker setup for a live multiparty interactive show.
Our far-field audio setup is required to be hands-free during live interaction and comprises four adjacent talkers with directional microphones in the same space. 
Such setups often introduce heavy cross-talk between channels, resulting in reduced automatic speech recognition (ASR) and natural language understanding (NLU) performance. 
To address this problem, we propose voice activity detection (VAD) model for all talkers using multichannel information, which is then used to filter audio for downstream tasks. 
We adopt a synthetic training data generation approach through playback and re-recording for such scenarios, simulating challenging speech overlap conditions. 
We train our models on this synthetic data and demonstrate that our approach outperforms single-channel VAD models and energy-based multi-channel VAD algorithm in various acoustic environments. 
In addition to VAD results, we also present multiparty ASR evaluation results to highlight the impact of using our VAD model for filtering audio in downstream tasks by significantly reducing the insertion error. 
\end{abstract}
\vspace{-3pt}
\begin{keywords}
Voice activity detection, multi-channel, cross-talk
\end{keywords}

\vspace{-3pt}
\section{Introduction}
\vspace{-3pt}
\label{sec:intro}

In natural group interactions with multiple talkers, overlapping speech segments occur frequently as a result of backchannels and interruptions~\cite{schegloff2000overlapping}. 
This is specifically evident in multiparty human-agent interactions as normal turn-taking protocols are often violated.
%An extreme version of this problem includes additional interfering background talkers and side conversations and is referred to as the cocktail party problem where there maybe additional interfering background talkers and side conversations.
%
Prior work in this direction has looked into structured scenarios such as meetings \cite{yu2022m2met, DiPCo}, 
and several datasets \cite{janin2003icsi,carletta2005ami} exist to facilitate research in multiparty conversations. 
Speaker diarization is a commonly studied problem and refers to the problem of identifying ``who spoke when", 
and the task of identifying speaker segments in a single audio channel that captures all talkers. 
Diarization is essential for downstream tasks such as Automatic Speech Recognition (ASR). 
As a result, several challenges \cite{watanabe2020chime, ryant2020third} have tried to improve the performance in this field by organizing benchmarked evaluations.
Traditional diarization approaches pre-process the audio stream using a speech activity detector (SAD), 
speaker segmentation and clustering using speaker dependent features \cite{park2022review}, 
whereas end-to-end formulations for diarization \cite{fujita2020end} that can be trained in a supervised manner 
are often better suited for online applications.
%
%Overlapping speech also poses challenges to speaker diarization and methods 
%such as overlap speech detection, speaker counting or Target Speaker appproaches \cite{cornell2022overlapped, medennikov2020target} 
%have emerged to deal with such multiparty scenarios.
%It must also be noted that single-channel based methods often have a larger model footprint and need to be more computationally intensive to obtain reliable features due to lack of information in the channel.
Additionally, multi-channel recordings are preferred in these applications with multiple talkers since they help emphasize the target speaker based on their proximity or direction \cite{zheng2022multi}.
% Furthermore, it might be possible to extract spatial features that leverage relation between multiple channels,
% when array geometry is already known or using geometry agnostic approaches \cite{raj2022gpu}. 
%Therefore, it is common practice to handle far-field multi-party situations by extracting robust features with simple front-end processing such as a beamformer.
%

In this work, we consider a multi-channel audio setup where each talker assigned a microphone channel is facing outwards from the center while talking for a live interactive show.
The constraints for our specific use-case arise from the design of a hands-free dialog-based multiparty human-agent interaction system where the agent response depends on knowledge of the active talker.
% We note that this layout is unusual when compared to a meeting scenario where talkers face each other or a display and their spoken interactions are easier to capture using a mic-array.
We install personal shotgun microphones at each talker position, for live interaction with the agent.
The recorded speech from microphone is fed to downstream ASR and NLU model for the interaction.
%  to avoid dependency on speaker characteristics that may be unreliable in far-field conditions. 
% Additionally, our model design is also motivated by low-latency requirements for online inference so that agent response times feel natural.
Although we employ directional microphones, each channel also picks up cross-talk from all other talkers depending on their proximity since the talkers are located in the same space.
Modern ASR models are trained to be robust to interfering background speech such as babble noise and show promising performance for recognizing primary talker's speech~\cite{whisper}.
However, with cross-talk even when the primary speaker is not talking, ASR models transcribe background speech due to their high sensitivity,
leading to insertion errors for ASR and degrading performance on downstream NLU tasks.
To resolve this problem, we need to reject cross-talk with low-latency by distinguishing foreground and background speech before sending the audio for downstream tasks such as ASR.

Conventional single channel VAD methods~\cite{sertsi2017robust, wagner2018deep,jia2021marblenet} represent poor performance when differentiating between near-field speech and cross-talk which shares characteristics with voiced speech.
Also, since our setup uses direcitonal microphone for outward speakers instead of microphone array, it is difficult to extract spatial information between microphones.
Prior attempts at multichannel VAD with multiple microphones have compared the energy ratio between different channels~\cite{ghosh2010robust,ichikawa21_interspeech}.
This may be helpful when finding the dominant near-field speaker but makes the assumption that the foreground talker is louder than background and hence not robust to variations in talker volume or microphone sensitivity.
Another candidate solution is speech separation, but these kinds of models are difficult to train on real-recorded datasets due to ambigous target.

In this paper, we propose a novel training framework for rejecting cross-talk by collecting real-world datasets and designing multiparty voice activity detector (\textbf{MPVAD}) to predict channel-wise speech activity using joint spectral features from all channels.
% We train a multiparty voice activity detector (\textbf{MPVAD}) to predict channel-wise speech activity using joint spectral features from all channels. 
This model has been motivated by several recent approaches that estimate acoustic parameters~\cite{yu2020room} or separate speech based on proximity~\cite{patterson22_interspeech}, 
which imply that near and far-field speech can be discriminated based on their spectral characteristics.
We hypothesize that by training on a joint prediction task the model implicitly learns to differentiate cross-talk from near-field speech.
To train the MPVAD robust to real environments, we also propose a synthetic data collection approach that simulate overlapping conditions for multiparty interaction scenarios.
Similar to LibriCSS~\cite{libricss}, we record dataset by playback of LibriSpeech on our setups and label target speech activity.

% We demonstrate our model's generalization ability on various real acoustic conditions 
We perform several experiments including application of downstream ASRs to verify that our model is able to differentiate between near-field and cross-talk speech, 
and hence can be used to detect speech activity from the primary talker in each channel in a speaker-agnostic way with low-latency.

% Our main contributions are as follows:
% \begin{itemize}
% \setlength\itemsep{-0em}
% \item A low-latency multi-channel speech activity model for active talker localization when using co-located microphones.
% \item A synthetic data collection approach similar to LibriCSS~\cite{libricss} to simulate overlapping conditions for multiparty interaction scenarios.
% \end{itemize}

\section{Multi-party multi-channel setup}
\label{sec:setup}

\begin{figure}[htb]

  \centering
  \centerline{\includegraphics[width=7.0cm]{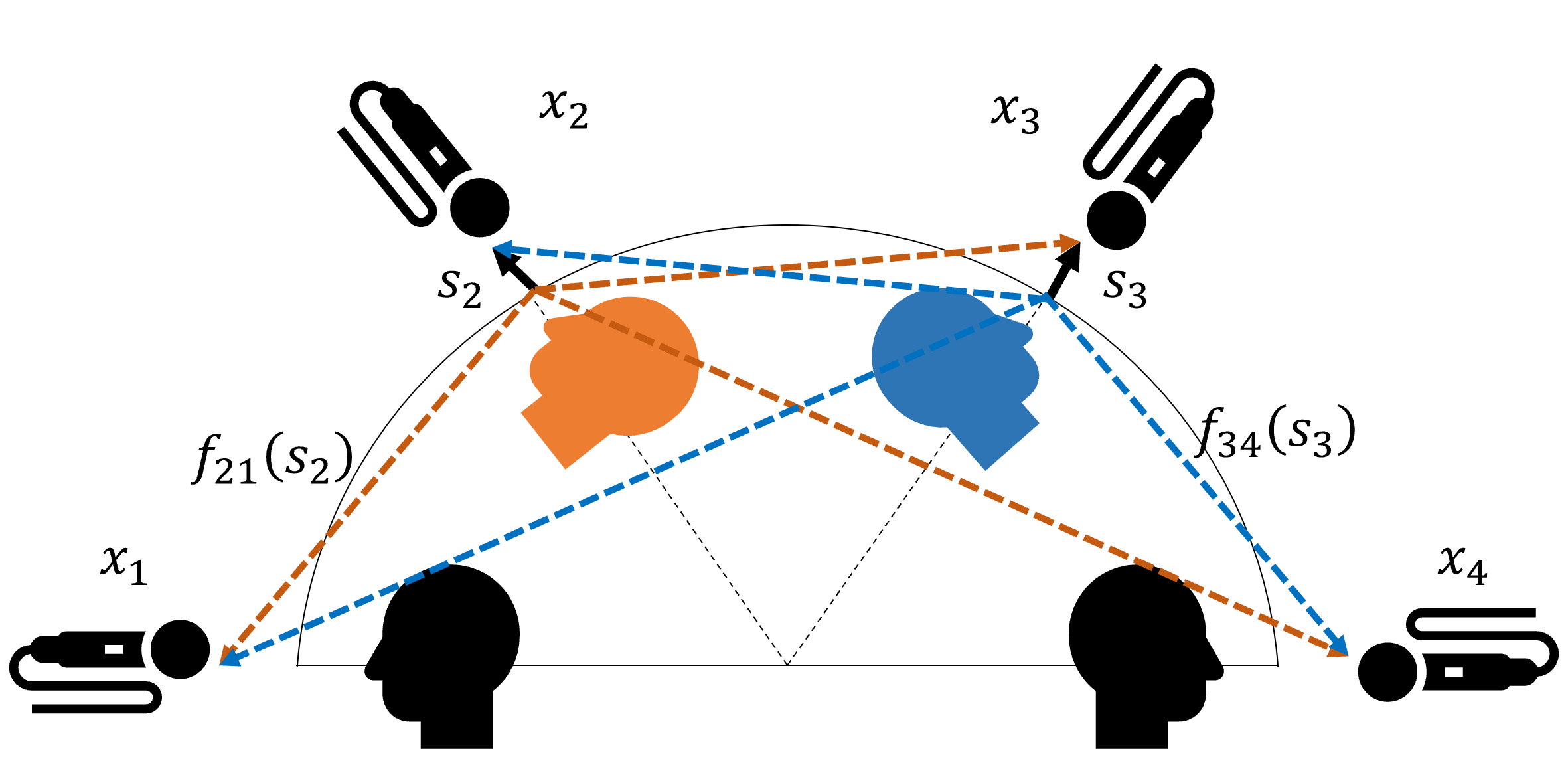}}
  \caption{Speaker and microphone layout used in our application and for the re-recording data collection described in Sec.~\ref{subsec:data}}
  \label{fig:layout}
\end{figure}

%TODO discuss different ways to solve multi-party and why we choose multichannel
Our multi-party interaction scenario consists of $4$ talkers facing outwards located roughly along the circumference of a semi-circle of diameter $\approx2.5$ m as shown in Fig.~\ref{fig:layout}. As mentioned earlier, each talker has a shotgun microphone pointed towards them at a distance $\approx1$ m away.
We assume that upto $4$ talkers can be active simultaneously and directional properties of shotgun mics provides certain degree of separation between different talkers.
% and that each talker is closer to their corresponding shotgun facing outwards from the center.
However, in practice due to cross-talk from adjacent talkers each microphone $i$ captures a mixture of all sources denoted by

$$
x_i =\sum_{j \neq i}  f_{ji}(s_j) + s_i~~\forall j\in\{1,2,3,4\}
$$
where $s_j$ and $f_{ji}$ indicates the speech from speaker $j$ and acoustic path from speaker $j$ to mic $i$ respectively.

Unlike \cite{libricss} which uses a single microphone array for re-recording,
we use individual shotgun mics on the periphery to capture talkers facing outwards from the center. 
While shotgun mics are directional and provide some degree of speaker isolation in each channel, our goal with the MPVAD model is to further enhance each talker's channel for downstream tasks such as ASR.
% We hypothesize that our model is able to use this speaker isolation to distinguish between a foreground and background talker.
%It must be noted though that in spite of using directional microphones, cross-talk between microphones still occurs in a far-field reverberant scenario.
We achieve this by building our multiparty VAD network $G$ to predict channel-wise speech activity using spectral features from multi-channel audio.

\section{Proposed Method}\label{sec:method}
% Shift 300ms is for real application but I think this description is not neccesary in the paper
We use a context window of $1000$ ms and predict channel activity for each second of multi-channel audio. 
In real-world applications, we run real-time MPVAD inference alongside streaming ASR models on all $4$ channels, running them in parallel. 
% Also this sentence correspond to real application but not our evaluation, so I'll remove that part
Once either channel's ASR returns an utterance transcription, the corresponding channel's VAD scores are looked up for that time interval to decide whether to retain the result for further processing.

% we predict channel activity for every one second of multi-channel audio~(offline processing).
% To consider contextual information and for practical deployment for ASR, 
% During inference for online combining with ASR process, it predicts activity every 300ms with 900ms of context.
% However, in this work we simply considered offline processing for evaluations.

\begin{figure}

\centering
\centerline{\includegraphics[width=6.5cm]{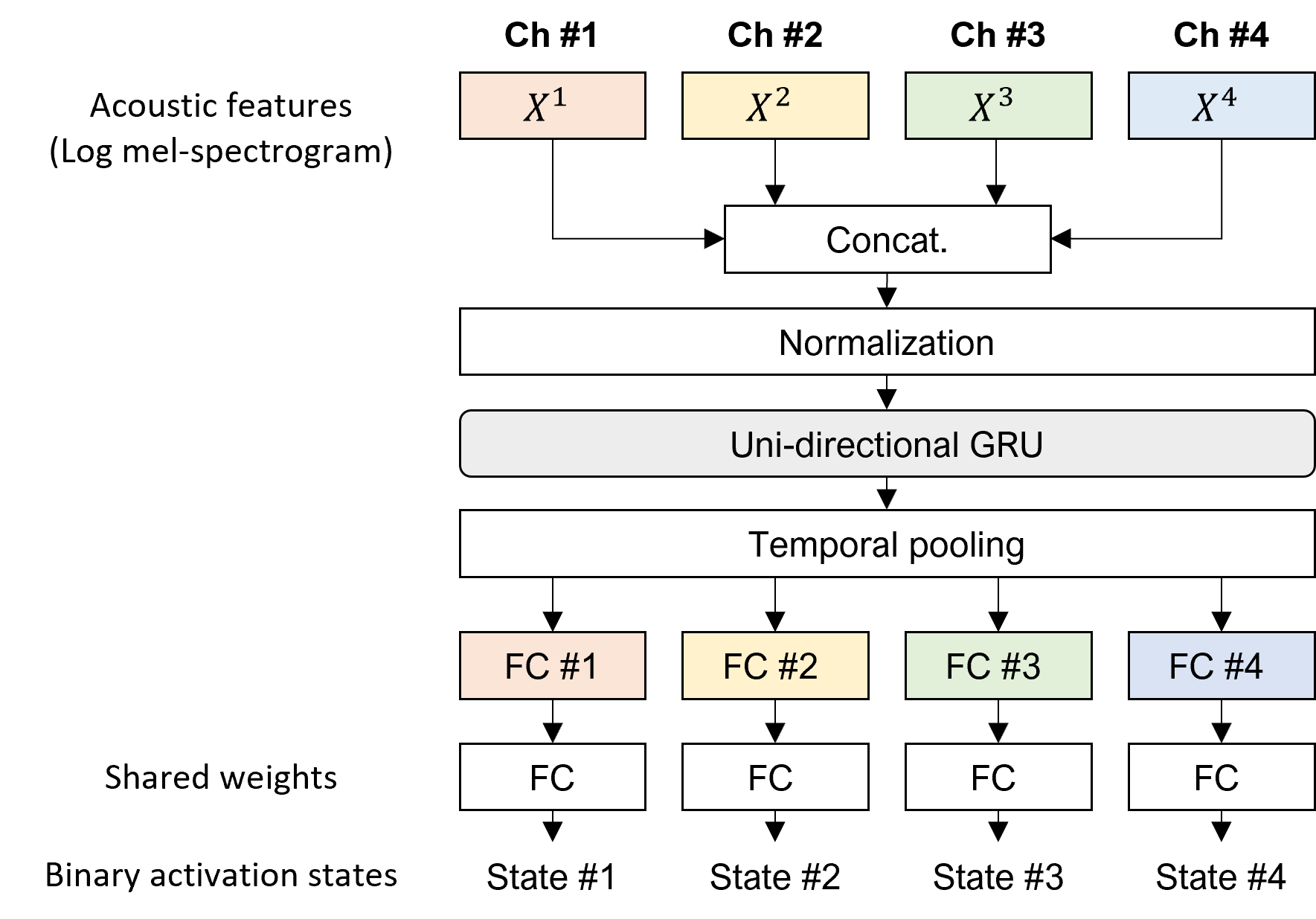}}
\caption{Network architecture of proposed MPVAD-MC model}\label{fig:proposed}
\end{figure}

Our MPVAD algorithm is described next.
Specifically, we compare two approaches: single-channel VAD models trained jointly focusing on modeling information from each channel, and a multi-channel VAD that can refer to all channels at once to model any relation between them. 
We demonstrate that both models learn complementary information and a posterior fusion of these methods adds to the robustness as discussed in Sec.~\ref{sec:result}. 
We refer to the fusion model that weights posterior of outputs from two models as {\bf MPVAD-F}.
%and fusion module that compute final outputs by voting estimation results from two modules.

The audio is first pre-processed by segmenting into 1s long windows followed by an amplitude normalization such that the resulting signal root mean square~(RMS) is $-25$ dB.
This normalization is employed so that the network can focus on acoustic characteristics such as harmonicity and reverberant condition rather than relying on energy alone. This makes the model invariant to external factors such as mic sensitivity or talker loudness.
We extract 40-dimensional log mel-spectrogram features as input to both our neural networks trained to output binary labels for each channel.
Further instance normalization is applied to input features to increase the stability of computation.

\noindent
{\bf Single-channel VAD~(MPVAD-SC)}
This model is designed to detect speech activity independently for each channel, and all model parameters are shared across channels.
% In this work, we use the same feature and network architecture for the single channel VADs to highlight the role of joint multichannel feature analysis.
The network architecture consists of a GRU layer to model temporal information among time frames.
%as shown in Fig.~\ref{fig:proposed}.
After GRU, temporal pooling is applied to extract segment-wise features which then serves as an input to the 2 fully-connected layers that output the final states.

\noindent
{\bf Multi-channel VAD~(MPVAD-MC)}
%We propose another subnetwork that detects channel activity based on inter-channel property.
Based on the observation that audio channels from adjacent microphones tend to have similar signals due to cross-talk, we propose a multi-channel model that
% Also, cosine similarity score between active channels have lower cosine similarity with high variance.
concatenates input features across all channels prior to the GRU layer (Fig.~\ref{fig:proposed}). This allows the model to exploit cross-channel information by learning a segment-level shared representation through temporal pooling.
%Feature extraction proceess is same as SCVAD and network architecture is also GRU.
This representation then goes into individual fully-connected (FC) layers to learn channel-specific information followed by a shared binary classifier for each channel.

\noindent
%{\bf Late fusion~(MPVAD-F)}
% fusion module outputs final prediction and 
%Although we applied separate FC layer for MCVAD, for generalization it tends to model inter-channel information.
%Otherwise, SCVAD tend to capture channel-specific information.
% Two kinds of network introduced above may have different characteristics (channel-specific, inter-channel relation).
%Therefore, to apply in real-world, trade-offs between two models should be compensated.
% we combine those models to produce final output.
% Inspired by the work mixture of experts~(MoE)~\cite{shazeer2017outrageously}, and doverlap~\cite{} which predict final outputs by voting to multiple predictions.
% Based on experiments, we found that two models have different roles, therefore, we combine those models to produce final output.
% We hypothesize that MCVAD model implicity models cross-talk conditions, and hence might not generalize well to all active or inactive talkers due to missing cross-talk.
% Hence the final output for each channel is computed by voting posterior probabilities from SCVAD and MCVAD (channel-specific feature and inter-channel feature) as follows:
% \begin{equation}
%  y_{mpvad}=(1-\alpha) \times y_{scvad} + \alpha \times y_{mcvad}, \alpha\in[0,1]
% \end{equation}
% where $\alpha$ is fixed weight ranging from 0 to 1.
% fusion network is simple fully-connected layer based on the overall confidence score~\cite{confnet} of similarity-based model output and spectrum-based output.

\noindent
% {\bf Training method}
%Based on the source speech activity, we can build speech activity for short length of segments.

Both MPVAD-SC and MPVAD-MC networks are trained using average binary-cross entropy loss on speech activity labels over 1s windows. 
The ground truth labels are derived from the source signals used for playback, and indicate when each channel's talker was active.
%TODO not sure what this means?
%To reduce the biased effect due to unbalanced dataset, we detect activity for each channel with binary states (foreground speech / others including cross-talk).
%Therefore, each module is learned by binary-cross entropy loss.
We also augment the dataset for MPVAD-MC network, by shuffling the order of channels during training, to prevent learning any permutation bias.
% And to learn fusion network, two models are frozen and only fusion network parameter is updated.

\vspace{-5pt}
\section{Experiment}
\label{sec:experiment}
\vspace{-5pt}
\begin{table}
    \centering
  % \Xline
  \begin{tabular}{c|c|c|c|c|c}
  \hline
  \multirow{2}{*}{Subsets}   & \multicolumn{5}{c}{Number of talkers} \\ 
  \cline{2-6}
  \multicolumn{1}{c|}{}  & 0~(sil.) & 1 & 2 & 3 & 4 \\ 
  \cline{1-6}
  % Need fixes for eval dataset
  Train & 4.87\% & 18.5\%  & 36.4\%  & 29.7\%  & 10.4\%\\ 
  Eval  & 9.75\% & 8.58\%  & 36.6\%   & 33.5\% & 11.5\% \\
  \hline
  \end{tabular}
  \caption{Statistics of recorded dataset by percentage of active talkers}
  \label{table:data_stats}
  \end{table}

\subsection{Synthetic data collection}
\label{subsec:data}
While datasets such as LibriCSS~\cite{libricss} exist for multi-talker multi-channel speech they lack certain properties needed for our application viz. a hands-free setup with talkers facing away from each other, personal mic channels and cross-talk between recorded channels.
To train and evaluate MPVAD models, we collect a dataset by playing back and re-recording utterances from Librispeech~\cite{panayotov2015librispeech} dev and test sets similar to \cite{libricss} but using a physical layout following our interaction scenario. 
%When constructing multi-channel utterances for playback we also design overlaps between talkers at random. 
In addition, since our final application uses Acoustic Echo Cancellation (AEC) for subtracting agent audio from talker channels, we record all audio channels post AEC to match expected acoustic conditions.

Following the layout shown in Fig.~\ref{fig:layout} we place $4$ Genelec speakers facing outwards along the semi-circle as our playback devices.
For each speaker, we create 20s long audio segments where each speaker is active only with a probability $p_{active}=0.6$. 
Silence is padded to utterances shorter than $20s$ to make all segments equal in length and simplify training from continuous audio recordings later. 
We use Sennheiser MKH 416-P48U3 shotgun mics for each channel and 
simultaneously playback and record on all $4$ output and input channels.
% Speakers for playback are placed with outward direction.
In addition to random chance of being active, we also assign a random target RMS energy for each segment from the set $\{-35, -25, -15\}$ dB following the energy levels used in the recipe for ~\cite{dubey2023icassp} and pre-normalize the audio to these levels for playback. 
This allows our model to be robust to energy variations expected in a real world scenario.
Overall, we record $1,000$ synchronized segments across 4 channels giving a total of 5.55 hrs. of audio simulating diverse cross-talk characteristics from multiple adjacent talkers.
Table~\ref{table:data_stats} shows the percentage of multi-talker overlap in our dataset.
When selecting utterances from Librispeech, we restrict a set of unique Speakers IDs for each talker position, and generate a playback list for each speaker by randomly sampling from these Speaker IDs and corresponding utterances.
From 1,000 segments, we use 950 for training and the other 50 for evaluation in this paper.

For training and evaluation of MPVAD models, we segment the audio recordings into 1 second windows using the shorter utterance length as a measure to extract exact overlapping segment boundaries. 
The resulting ratio of these windows by the number of active speakers are represented in Table.~\ref{table:data_stats}.
When evaluating impact on ASR, full segments are decoded to prevent continuity errors.
To verify how our model generalizes, we also collected similar datasets using slightly different talker layouts ~(noted as \textbf{Set B} and \textbf{Set C}).
These two sets are also recorded in a different space compared to \textbf{Set A}, and additionally talkers in \textbf{Set C} are arranged along vertices of a square.
We train the MPVAD network on audio from \textbf{Set A} and present cross-validation results on \textbf{Set B} and \textbf{Set C}.
%\textbf{Set B} is recorded in different room compared to \textbf{Set A}, and furthermore, \textbf{Set B} is recorded with rectangular layout. 
% The audio is recorded at a sampling rate of $44.1 kHz$, and then resampled to $16 kHz$ for processing to reduce computational overhead.
% For evaluation of speech recognition performance after processing with channel activity detection, 
The recorded audio segments are resampled to $16$ kHz before training.

\subsection{Network and training configuration}
\label{ssec:config}
We extract 40 dimensional log-scale mel-spectrogram features using a Short-time Fourier-transform~(STFT) with a 512-point FFT. The Fourier transforms are computed over 20 ms windows and a 10 ms hop length using a Hamming window. 
% We use torchaudio's feature extractors for this purpose for efficient feature extraction on GPUs.
%Our network is trained over 1s long windows to provide long term context for the classifier.
MPVAD-SC and MPVAD-MC use 64 and 16 dimensional GRU layers respectively, and 16 dimensional FC layers after GRU.
Hyperbolic tangent was used as nonlinear function between FC layers.
%Weight for fusion was set to $\alpha=0.75$ after hyperparameter optimization on a validation set.
Both networks were trained with AdamW optimizer with initial learning rate 0.001.
\vspace{-4pt}
% The proposed model achieves a real time factor~(RTF) of 150x on a single A100 GPU and 15x on CPU, which demonstrates its suitability for the streaming ASR scenario.

\section{Results and analysis}
\label{sec:result}

We evaluate the second-level prediction accuracy of the proposed multiparty VAD model over 1 second frames against VAD ground-truth and ASR performance after speech enhancement based on cross-talk estimation for each channel.
We also compare against other single-channel public VAD models such as WebRTCVAD~\footnote{https://github.com/wiseman/py-webrtcvad}, and Silero~\cite{SileroVAD}. We use the pretrained models for both these systems and aggregate the voice activity predictions for each $10-30$ms over 1 second windows to enable a direct comparison with our proposed approach.
%Since these are not originally designed for multi-channel scenario, we run them on all channels independently to highly issues arising from lack of robustness to cross talk.

We also adopt the energy-based approach proposed in \cite{ichikawa21_interspeech} as a baseline, since they consider a multichannel scenario similar to ours for cross-talk conditions. 
We implement an energy based MVAD model analogous to \cite{ichikawa21_interspeech} by training SVM models to predict activity per channel. We use ambient noise as reference to normalize RMS energy in each channel and use them as a multichannel feature for each frame.
Furthermore, we compare both the proposed models:  MPVAD-SC, MPVAD-MC and a late posterior fusion MPVAD-F of the two models.

% Final model
% SCVAD:
% MCVAD:
% For evaluation of speech recognition performance after processing with channel activity detection, 

%===============================Bar graph===================================
% TODO: compute error bar (not filled now)
\pgfplotstableread[row sep=\\,col sep=&]{
    interval & MPVAD-SC & MPVAD-MC & MPVAD-F & e_sc & e_mc & e_mp\\
    % 0  & 99.33 & 92.50 & 96.04 & 1.88 & 2.08 & 2.04 & 2.03\\
    % 1  & 76.89 & 98.18 & 97.17 & 1.71 & 1.4 & 1.38 & 1.36\\
    % 2  & 77.22 & 97.18 & 95.64 & 1.4 & 1.04 & 0.98 & 0.97\\
    % 3  & 85.08 & 95.02 & 95.36 & 1.88 & 2.08 & 2.04 & 2.03\\
    % 4  & 96.19 & 89.60 & 92.03 & 1.71 & 1.4 & 1.38 & 1.36\\
    0  & 99.3 & 92.5 & 96.2 & 0.65 & 0.43 & 0.31\\
    1  & 76.9 & 98.2 & 97.0 & 3.61 & 0.23 & 0.3\\
    2  & 77.2 & 97.2 & 95.9 & 1.73 & 0.14 & 0.17\\
    3  & 85.1 & 95.0 & 95.4 & 1.54 & 0.19 & 0.18\\
    4  & 96.2 & 89.6 & 92.4 & 1.41 & 0.46 & 0.4\\
    }\mydata

%\vspace{-2pt}
%\begin{figure*}
%    \begin{minipage}[h]{1.25\columnwidth}
    
\begin{figure}
    \begin{tikzpicture}
        \centering
            \begin{axis}[
                    width=1.\columnwidth, height=4cm,
                    scaled y ticks = false,
                    scaled x ticks = false,
                    xtick pos=left,
                    ytick pos=left,
                    ybar=5pt,
                    bar width=6pt,
                    enlargelimits=0.15,
                    ylabel={Selection ratio (\%)},
                    minor tick length=1ex,
                    major x tick style = {opacity=1},
                    ymajorgrids=true,
                    grid style=dashed,
                    legend style={at={(.72, 1.25)}, font=\footnotesize,
                    anchor=north, legend columns=-1},
                    legend image code/.code={
                            \draw [#1] (0cm,-0.1cm) rectangle (0.12cm,0.20cm); },
                    % % enlargelimits=0.15,
                    % ybar interval=0.7,
                    % ybar=6pt,
                    % legend style={at={(0.5,-0.3)},
                    %     anchor=north,legend columns=-1},
                    symbolic x coords={0, 1, 2, 3, 4},
                    xtick=data,
                    nodes near coords,
                    nodes near coords align={vertical},
                    % every node near coord/.append style={font=\footnotesize},
                    every node near coord/.append style={font=\footnotesize},
                    ymin=75,ymax=100,
                    ylabel={Accuracy (\%)},
                    xlabel={Number of talkers},
                    every axis y label/.style={at={(ticklabel cs:0.5)},rotate=90,anchor=near ticklabel},
                ]
                % \addplot [style={black,fill=gray}, error bars/.cd, y dir=both, y explicit] table[x=interval,y=SCVAD,y error=e_sc]{\mydata};
                % \addplot [style={blue,fill=blue!30!white}, error bars/.cd, y dir=both, y explicit] table[x=interval,y=MCVAD,y error=e_mc]{\mydata};
                % % \addplot [style={blue,fill=blue!30!white}, error bars/.cd, y dir=both, y explicit] table[x=interval,y=FB,y error=e_fb]{\mydata};
                % \addplot [style={red,fill=red!30!white}, error bars/.cd, y dir=both, y explicit] table[x=interval,y=MPVAD, y error=e_mp]{\mydata};
                \addplot [style={black,fill=gray}] table[x=interval,y=MPVAD-SC]{\mydata};
                \addplot [style={blue,fill=blue!30!white}] table[x=interval,y=MPVAD-MC]{\mydata};
                \addplot [style={red,fill=red!30!white}] table[x=interval,y=MPVAD-F]{\mydata};
                ;

                % \legend{Unprocessed, Backbone, Proposed}
                % \legend{B, FB, Proposed}
                \legend{MPVAD-SC, MPVAD-MC, MPVAD-F}
                % \legend{Unprocessed, B, Proposed}
            \end{axis}
        \end{tikzpicture}
    % \end{minipage}

    \caption{Channel activity detection accuracy vs. number of talkers (over the entire evaluation set).
    }
    \label{fig:ser_graph}
    %    \end{minipage}
\end{figure}
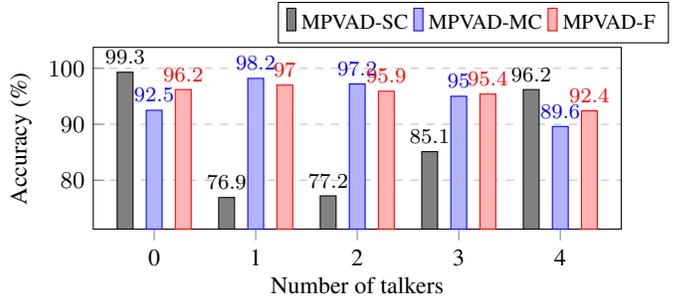
    %\hfill
    %\begin{minipage}{0.75 \columnwidth}
        % \begin{table}[]
            % \footnotesize
        %   \centering
        % \Xline

 \begin{table}
        \centering
        \begin{tabular}{c|c|c|c}
        \hline
        \multirow{2}{*}{Methods}  & \multicolumn{3}{c}{Eval sets} \\ 
        \cline{2-4}
          & A & B & C \\ 
        \cline{1-4}
        WebRTC~(Soft)    & 64.77& 61.94& 62.82\\ 
        WebRTC~(Hard)    & 51.95& 48.39& 47.66\\ 
        Silero~\cite{SileroVAD}     & 77.49& 63.40& 61.65\\
        Energy~\cite{ichikawa21_interspeech} & 96.92 & 58.80  & 56.64  \\ \hline
        % SCVAD~(small) & \textbf{100.0} & 96.02  &  96.01  & 93.75 & 94.92 & 95.52\\ 
        MPVAD-SC & 96.25 & 78.17 & 78.07\\ 
        % MCVAD~(Cosine) & xx & xx  & xx   & xx & xx\\
        MPVAD-MC & 98.65 & 93.14 & 93.85\\
       % MPVAD~($\alpha=0.25$) &97.96 & 83.68 & 83.59\\
       % MPVAD~($\alpha=0.5$) &98.64 & 89.57 & 90.02\\
        MPVAD-F~($\alpha=0.75$) &\textbf{98.82} & \textbf{93.24} & \textbf{94.27}\\
        \hline
        \end{tabular}
        \caption{Multi-channel voice activity detection depending on different acoustic conditions. 
        $\alpha=0.75$ fusion indicates the weight given to MPVAD-MC after tuning}
        \label{table:acc}
\end{table}
    %\end{minipage}
%\end{figure*}

\subsection{Channel activity detection performance}
% \noindent \textbf{Channel activity detection}
% \noindent \textbf{Comparison with baselines}
Table~\ref{table:acc} shows accuracy for the proposed MPVAD models and baseline methods at the frame-level on evaluation \seta, \setb~and \setc.
We note that conventional VAD methods demonstrate a higher error due to ambiguity between acoustic characteristics of background and foreground speech.
For WebRTCVAD, we compare two different thresholds by switching the mode between 0~(soft) and 3~(hard). 
We find that aggresive thresholding (hard) is good for rejecting cross-talk, but leads to a lot of misses with target close-talk not being detected as active depending on speech energy. On the other hand, the less aggresive soft mode had fewer misses but a lot of false-alarms, leading to an overall poor result. 
Silero\cite{SileroVAD} showed better performance than WebRTCVAD, but was still not able to reject cross-talk.
Our energy based MVAD baseline inspired from \cite{ichikawa21_interspeech} outperforms all other baseline methods, 
as it was trained on \seta, but fails to generalize to \setb, and \setc. 
We hypothesize that this might be due to different reverberant conditions in the two rooms and hence energy levels on which the models are trained. 

Finally, both our proposed models show robust performance across evaluation \seta. 
MPVAD-SC however generalizes poorly to evaluation sets: \setb~and \setc~indicating that it might also be relying on energy characteristics of each channel. 
MPVAD-MC on the other hand focuses on cross-channel features and is hence more robust to change in acoustic environment and channel conditions. 
%Comparing to VAD model processing single-channel, VAD with multi-channel input outperforms for generalizing in different room environments.
% Single-channel VAD trained in training set A dataset showed poor generalization performance among different room environment.
% We can conclude that setting simple is not good for single-channel feature when geometry and room environmen.
% We could verify that contribution of feature was different depending on number of channels.

\begin{figure}[htb]

  \begin{minipage}[b]{0.3\linewidth}
    \centering
    \centerline{\includegraphics[width=3cm]{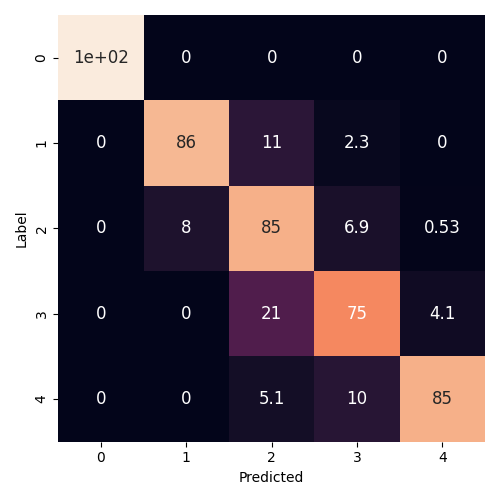}}
  %  \vspace{1.5cm}
    \centerline{(a) MPVAD-SC}\medskip
  \end{minipage}
  \hfill
  \begin{minipage}[b]{0.3\linewidth}
    \centering
    \centerline{\includegraphics[width=3cm]{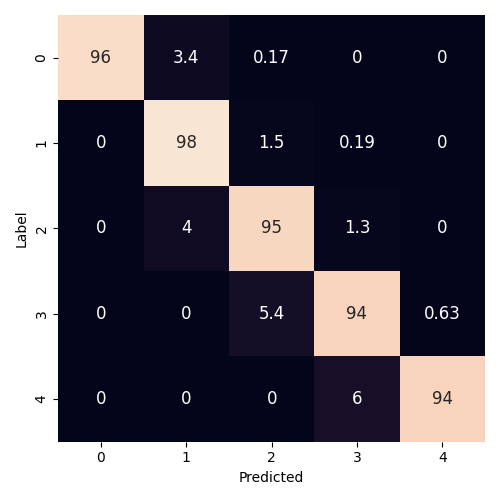}}
  %  \vspace{1.5cm}
    \centerline{(b) MPVAD-MC}\medskip
  \end{minipage}
  \hfill
  \begin{minipage}[b]{0.3\linewidth}
    \centering
    \centerline{\includegraphics[width=3cm]{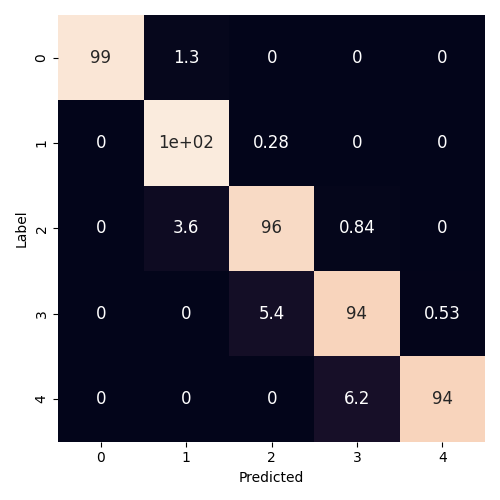}}
  %  \vspace{1.5cm}
    \centerline{(c) MPVAD-F}\medskip
  \end{minipage}

  \begin{minipage}[b]{0.3\linewidth}
    \centering
    \centerline{\includegraphics[width=3cm]{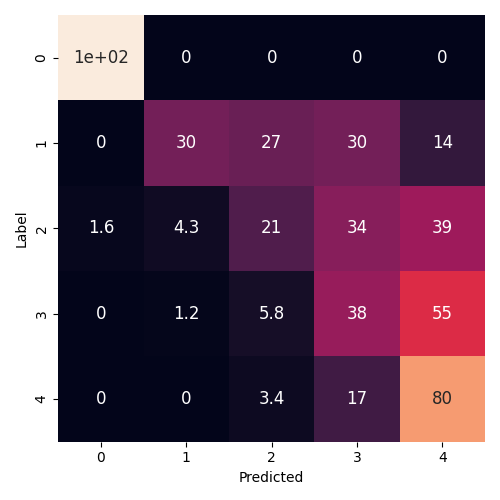}}
  %  \vspace{1.5cm}
    \centerline{(d) MPVAD-SC}\medskip
  \end{minipage}
  \hfill
  \begin{minipage}[b]{0.3\linewidth}
    \centering
    \centerline{\includegraphics[width=3cm]{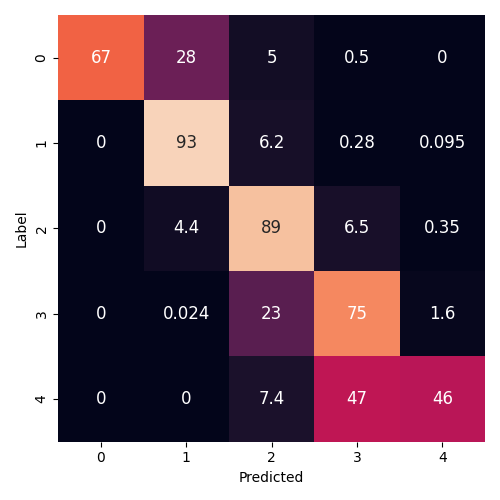}}
  %  \vspace{1.5cm}
    \centerline{(e) MPVAD-MC}\medskip
  \end{minipage}
  \hfill
  \begin{minipage}[b]{0.3\linewidth}
    \centering
    \centerline{\includegraphics[width=3cm]{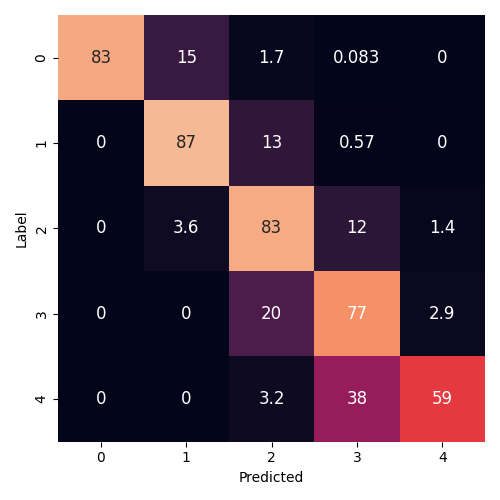}}
  %  \vspace{1.5cm}
    \centerline{(f) MPVAD-F}\medskip
  \end{minipage}

\caption{Confusion matrices showing how different models perform on the speaker counting task. The true number of active speakers are presented along rows and columns show estimated number of speakers based on VAD predictions. (a)-(c): Evaluation set A (d)-(f): Evaluation set C}
\label{fig:confusion}
\end{figure}
\vspace{-0.5cm}
\subsection{Ablation study: performance vs. number of active talkers}

We also perform error analysis to investigate model performance for different number of active talkers. 
This is presented in Fig.~\ref{fig:ser_graph} and shows the difference in performance for single-channel vs. multi-channel models.
To further understand model behavior, we count number of active speakers based on channel-wise MPVAD predictions.
Fig.~\ref{fig:confusion} shows that MPVAD-MC is more robust than MPVAD-SC in the presence of cross-talk, even under different room environment and channel conditions.
However, when applied to a different environment, it performs poorly in cases with all active or all silent talkers. 
This might be the case because it relies on multi-channel features and cross-talk like characteristics for learning channel activity information.
We observe that MPVAD-SC on the other hand performs better in these cases since it only relies on channel-specific characteristics, but tends to confuse cross-talk with foreground speech more often. We also observe empirically that fusion of the two provides a slightly better and balanced prediction (MPVAD-F), however in practice there isn't a noticeable difference when deploying MPVAD-MC in our live interaction scenario. The MPVAD-MC model generalizes well to live conversational speech from unseen speakers.

%We hypothesize that by fusing output from these 2 systems, MPVAD might be able to leverage complementary information from each. We demonstrante this via late fusion via convex weighting of the posteriors where $\alpha$ is the weight applied to MCVAD prediction and $(1-\alpha)$ is applied to the SCVAD prediction.
%In case of different sets, it did not achieve best performance for specific number of talkers, but it also filled the gap between two models in overall.

\vspace{-0.2cm}
\subsection{Impact on ASR performance}
% \noindent \textbf{Impact on ASR}
To understand the impact of the system on downstream ASR, we propose a simple enhancement strategy by masking the speech signal based on estimated voice activity detection for each channel. Each channel is processed by cross-talk suppression followed by ASR decoding.
% For ASR, we used xx model from Kaldi toolkit~\footnote{}.
We used the pre-trained Whisper model~($base\_en$)\footnote{https://github.com/openai/whisper}~\cite{whisper} in this study.
% This models were run in parallel to recognize speech from each channel.
Channel-wise Word Error rate (WER) was computed based on the original transcription using the NIST Sclite tool. 
Recognition results are demonstrated in Table~\ref{table:wer_result}, in terms of insertion, deletion, and substitution errors.
We observe that ASR results shows high insertion errors due to high sensitivity to background talkers, when compared against using oracle VAD i.e. segment boundaries based on known utterance length.
Since conventional VAD methods~(WebRTCVAD, Silero) are not robust to cross-talk, they reduce insertion errors but also increase deletion and substitution errors due to undesired distortion of speech.
Our method effectively captures close-talk characteristics and helps greatly reduce the insertion errors by properly rejecting cross-talk from other speakers.
This evaluation in addition to our live interaction tests demonstrate that the proposed multi-channel VAD is effective for improving ASR performance in our multi-talker interaction scenario.
The proposed model achieves a real-time factor~(computed as inference time divided by processed audio length) of 0.0067 on a single A100 GPU 
and 0.0667 on AMD EPYC 7763 64-Core CPU, which demonstrates its suitability for the streaming ASR scenario.
\begin{table}[h]
  \centering
% \Xline
\begin{tabular}{c|c|c|c|c}
\hline
\multirow{2}{*}{Methods}   & \multicolumn{4}{c}{Error category} \\ 
\cline{2-5}
  & Ins. & Del. & Sub. & Total \\ 
\cline{1-5}
Unprocessed     & 63.0 &  1.3  & 6.1  & 70.4\\ % Done!
Oracle seg    & 5.7  & \textbf{1.2}   & 6.2 & 13.1 \\ % TODO
\hline
% TODO: Compute on webrtcvad with mode 0
WebRTC~(Soft)    & 53.3  &  3.5  & 6.9  & 63.7 \\ % Done!
WebRTC~(Hard)    & 20.6  &  15.6  & 8.9  & 45.1 \\ % Done!
Silero~\cite{SileroVAD}    & 12.2  & 17.0   & 10.1 & 39.3\\ \hline
% MCVAD~\cite{ichikawa21_interspeech}    & xx  & xx   & xx & xx \\\hline
% TODO: Compute on MPVAD with final version
\textbf{MPVAD-F}  & \textbf{2.4}   & \textbf{1.2} & \textbf{5.9} & \textbf{9.5}\\
\hline
\end{tabular}
\caption{WER~(\%) result with pre-trained Whisper ASR evaluated on set A.}
\label{table:wer_result}
\end{table}

\vspace{-1cm}

\section{Conclusion}\label{sec:conclusion}
% End
In this work, we compared different approaches to multi-party voice activity detection with simultaneous talkers using multiple far-field personal microphones. We develop a low-latency approach for \mbox{MPVAD} that takes into account cross-talk between channels and 
show that our algorithm learns to efficiently discriminate between acoustic characteristics from foreground and background talkers when trained jointly using features from all channels. We also demonstrate its impact in reducing insertion errors for downstream ASR tasks by suppressing audio during intervals where cross-talk is predicted.

For future work, we would like to extend our multi-party VAD method to channel-identity agnostic approaches such as target-speaker VAD such that the constraint due to personal mics can be removed.
We also plan to experiment with fusion of multimodal information and spatial audio features which should be beneficial for tasks involving overlap and speaker change detection\cite{mariotte23_interspeech}. 
In addition, our training data generation could also make use of more sophisticated overlapped speech generators ~\cite{park2023property}.
% References should be produced using the bibtex program from suitable
% BiBTeX files (here: strings, refs, manuals). The IEEEbib.bst bibliography
% style file from IEEE produces unsorted bibliography list.
\vfill\pagebreak
% -------------------------------------------------------------------------
\bibliographystyle{IEEEbib}
\bibliography{strings,refs}

\end{document}